\documentclass[12pt,a4paper]{article}

\usepackage{amsfonts,amsmath,amssymb,bm}
\usepackage{graphicx,setspace}
\title{\vspace{-48pt}Accuracy of the box-counting algorithm\\ for noisy fractals}

\author{A. Z. G\'orski$^{a}$, M. Str\'o\.z$^{a,b}$, P. O\'swi\c ecimka$^{a}$, J. Skrzat$^{c}$\and%
$^{a}$H. Niewodnicza\'nski Inst. of Nucl. Physics, Polish Acad. of Sci.,\\% 
Radzikowskiego 152, Krak\'ow, 31-342%
\and % 
$^{b}$AGH University of Science and Technology, Faculty of Physics\\% 
and Applied Computer Sci., Krak\'ow%
\and % 
$^{c}$Dept. of Anatomy, Collegium Medicum, Jagellonian Univ., Krak\'ow\\}

\date{}

\begin{document}

\maketitle

\begin{abstract}
The box-counting (BC) algorithm is applied to calculate fractal dimensions 
of four fractal sets. The sets are contaminated with an additive noise 
with amplitude $\gamma = 10^{-5} \div 10^{-1}$. 
The accuracy of calculated numerical values of the fractal dimensions 
is analyzed as a function of $\gamma$ for different sizes 
of the data sample ($n_{tot}$). 
In particular, it has been found that a tiny ($10^{-5}$) addition of noise 
generates much larger (three orders of magnitude) error of the calculated 
fractal exponents. 
Natural saturation of the error for larger noise values prohibits
the power-like scaling. Moreover, the noise effect cannot be cured 
by taking larger data samples.

\end{abstract}

\section{Introduction}

 In last decades computations of fractal dimensions (exponents)
have become very popular. 
A search through the Web of Science reveals about $10,000$ journal articles
with the term {\it fractal dimension} in title, keywords, or abstract 
\cite{Gnaiting03}.
The fractal structures have been found in a wide spectrum of problems, 
ranging from high energy physics \cite{Bialas} to cosmology \cite{Chmaj} and from 
bio-medical sciences \cite{Skrzat} to econophysics \cite{AZG2002}.

Although the concept of fractal exponents has been popularized long time ago
\cite{Mandelbrot77,Theiler}, the accuracy of obtained results is usually either 
not discussed or overestimated. Moreover, it has been found that in quite 
a few papers misleading numerical results and conclusions have been published
(see {\it e.g.}: \cite{Skrzat,AZG2001,McCauley02,Castiglioni10}). 
In many cases, especially for the time series analysis, the fractal exponents are 
computed indirectly, from the Hurst exponents. However, a simple relation between 
the two exponents holds only in some special cases \cite{Jaffard97}.
What is more important, the computational algorithms of fractal exponents
have systematic errors much larger than the corresponding statistical 
errors. This is mainly due to the finitness of the data sample (of size 
denoted by $n_{tot}$). This problem is independent on the 
computational algorithm applied. For the box-counting (BC) algorithm, 
it has been shown that the error (roughly)
scales according to the power law, $\sim 1/n_{tot}^\alpha$ \cite{AZGczI},
for pure fractals, without noise.

 The aim of this paper is to perform an analysis of the fractal structures contaminated 
with an additive external noise and to estimate the accuracy of the BC algorithm
in these cases.
A similar analysis for the influence of noise on the Hurst exponents, which were
computed by the MF-DFA method, has been performed recently \cite{Ludescher11}. 
It should be stressed that the type of algorithm is not so important, 
as the accuracy depends more on the choice of a particular fractal set than 
on the type of the algorithm applied. Algorithms that work better 
with some classes of the fractal sets are not so efficient for other classes
(see {\it e.g.} Ref. \cite{Brewer06}).

We generate the mathematical fractal sets of different size ($n_{tot}$), 
namely, the standard (1/3) Cantor set (CS), 1/2-Cantor set (in 1-dimensional
embedding space), Sierpi\'nski triangle and Sierpi\'nski carpet
(in 2-dimensional embedding space). 
The data set $\{x_i\}$ is contaminated by the external additive noise,
$\xi_i$ ($\vert\xi_i\vert\le 1$), with different amplitudes $\gamma$
\begin{equation}
x_i \rightarrow x_i + \gamma \ \xi_i \ .
\label{noisedef}
\end{equation}
For $\gamma=0$, no external noise is present. The values of $\gamma$ 
that will be considered are in the range $10^{-5} \div 10^{-1}$
(four orders of magnitude). 
Also, the size of the data sample, $n_{tot}$, has been varied in the range
$2^6 \div 2^{20}$ (about four orders of magnitude). 
The fractal exponent has been calculated using the BC algorithm, 
according to the standard formula \cite{Mandelbrot77}
\begin{equation}
d(q) = \frac{1}{1-q} \, \lim_{N\to\infty} 
  \frac{\log  \sum_i  p_i^q(N)}{\log N} \ , 
\label{fracdimdef}
\end{equation}
where $N$ denotes the number of divisions, $p_i(N)$ is the measure of the subset
in the $i$th box for a given division $N$ and the box size $\epsilon = 1/N$. 
In practical calculations the numerical data sets are always finite. 
Hence, the scaling range (the limit $N\to\infty$) is chosen as the interval
within which the log-log plot is closest to the linear function. 
In fact, the problems with choosing the proper scaling range are the main sources 
of the systematic error. 
Furthermore, for monofractal sets one has $d(q)= d_0$ = const.
All cases discussed here are monofractals and one can set $q=0$.

  In the following section, the numerical results will be 
presented and discussed. The final section is devoted to summary 
and conclusions.

\section{Accuracy analysis}

%%%%%%%%%%%%%%%%%%%%%%%%%fig%1%%%%%%%%%%%%%%%%%%%%%%%%%%%%%%%%%%%%%%%%%%%%%%%
\begin{figure}
\begin{center}
 \includegraphics[width=12.0cm,angle=0]{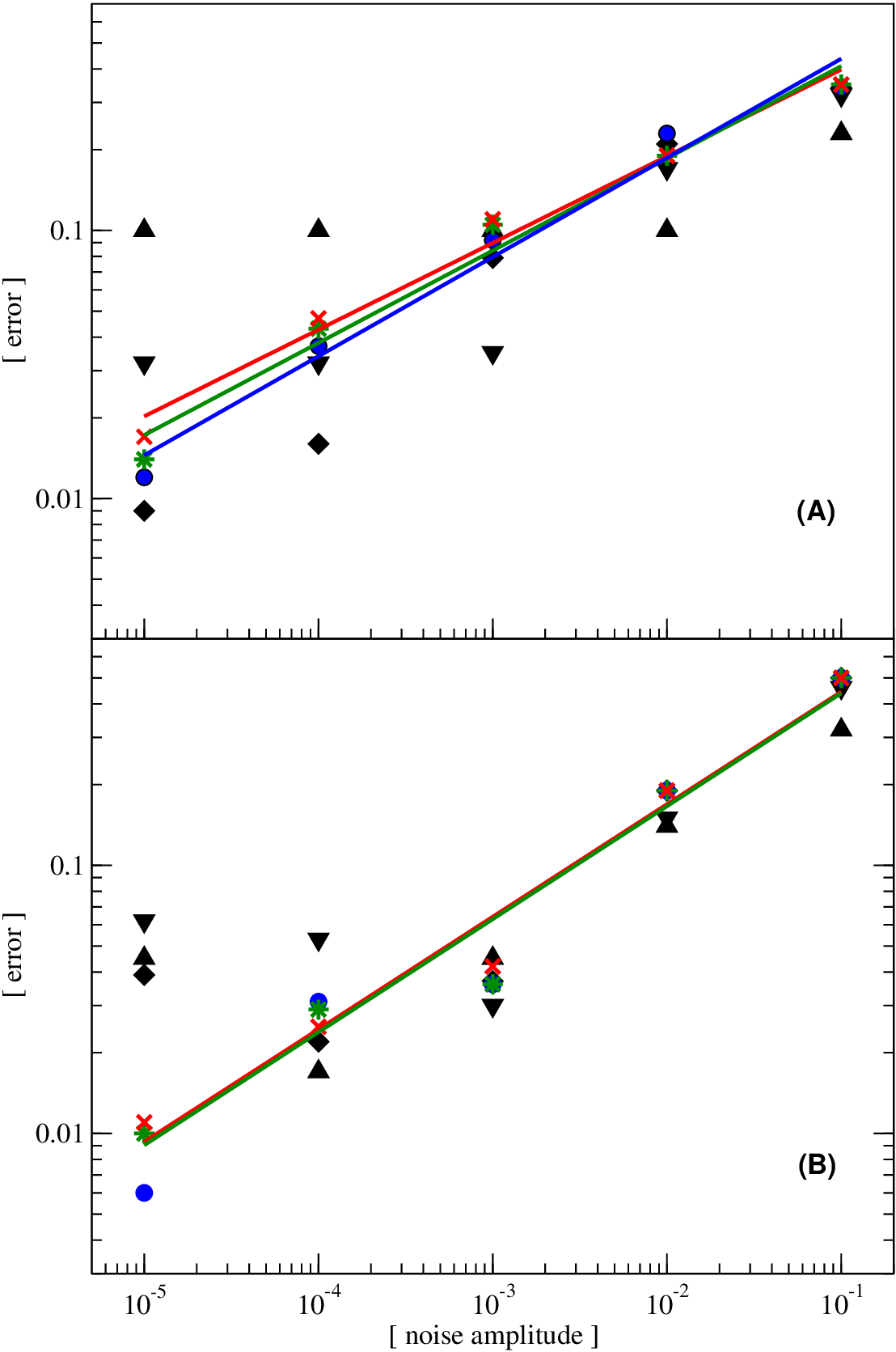}
 \caption{Error $\Delta d$ as a function of the noise amplitude ($\gamma$)
for the standard Cantor set (A) and for the 1/2-Cantor set (B) with the different 
sizes of the data sample, $n_{tot} = 2^{6}, 2^{8}, 2^{10}, 2^{12}, 2^{14}, 2^{16}$.
The errors are denoted by the triangle-up, triangle-down, diamond, circle, star 
and cross symbols, respectively. 
In the last cases, the power-like fits are also added (solid lines).}
\end{center}
\end{figure}

 For our analysis, the well-known fractals embedded in the 
one- and two-dimensional embedding space are taken, for which 
the precise theoretical values of thefractal exponents ($d^{th}$) 
are well known. 
Approximate values of these exponents ($d$) are calculated with the BC algorithm,
according to Eq. (2), and the corresponding error is determined as follows
\begin{equation}
 \Delta d = \vert d^{th} - d \vert \ .
\label{errordef}
\end{equation}
These computation is repeated for different values of the noise amplitude 
$\gamma$ and for different values of the sample size $n_{tot}$. 
In effect, one can find a dependence of the error on the noise strength 
and the sample size. 
Because $\Delta d \le E$, from Eq. (\ref{errordef}), it is clear that 
the error $\Delta d$ has the following upper bound
\begin{equation}
 \Delta d \le E - d^{th} \ \ ,
\label{errorbound}
\end{equation}
where $E$ is the embedding dimension. Hence, the error can be at most 
of order of $1$. The question remains, how fast the bound 
(\ref{errorbound}) is reached. This will be studied below.

 First,  the standard Cantor set and the 1/2-Cantor set were analyzed. 
The noise dependence of the errors was shown in Figs.~1-A and 1-B, respectively. 
Here, the different sample sizes are represented by the triangles-up, triangles-down, 
diamonds, circles, stars and crosses, respectively. For larger data sets,
the power fits were included for a comparison.

%%%%%%%%%%%%%%%%%%%%%%%%%fig%2%%%%%%%%%%%%%%%%%%%%%%%%%%%%%%%%%%%%%%%%%%%%%%%
\begin{figure}
\begin{center}
 \includegraphics[width=12.0cm,angle=0]{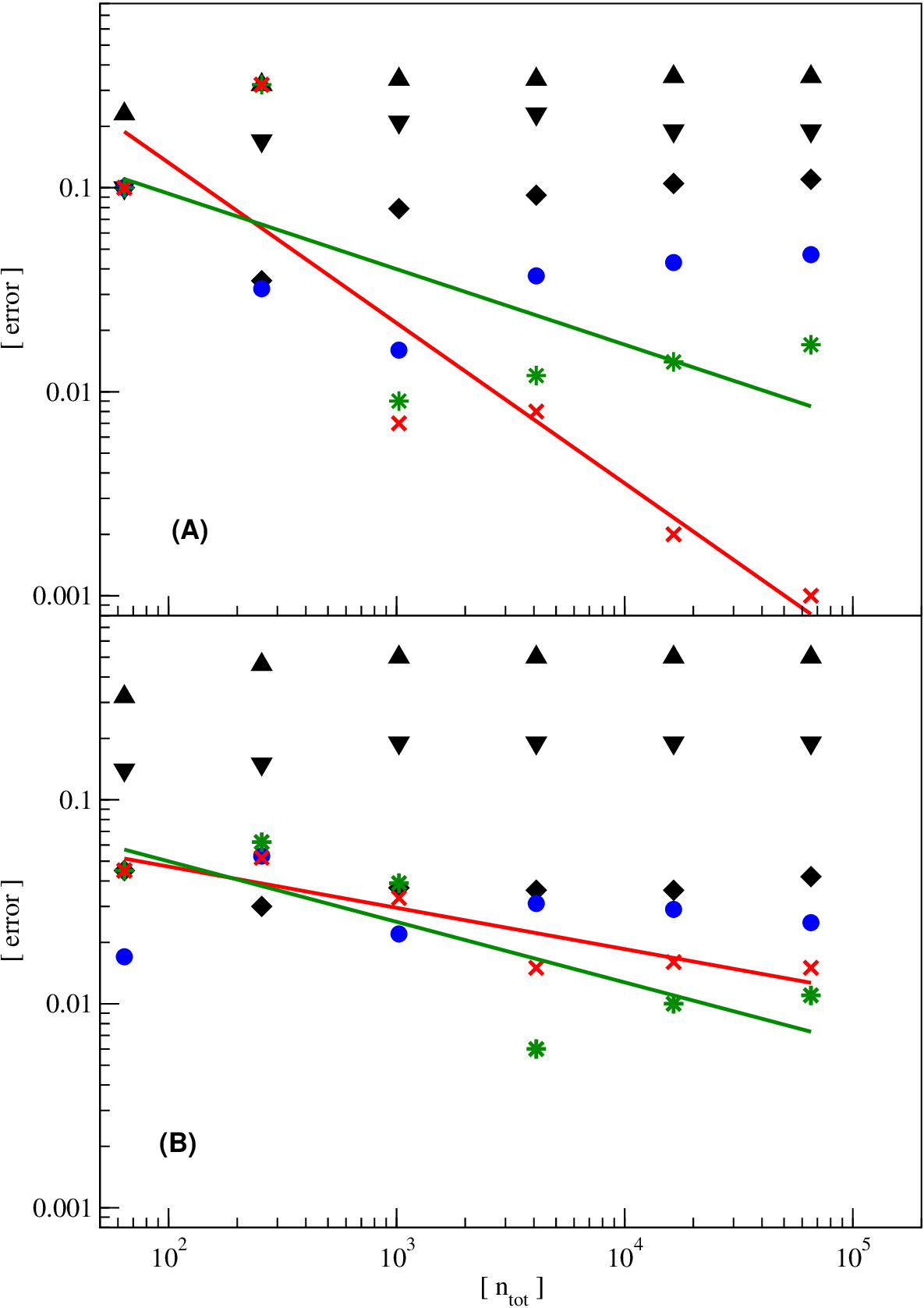}
 \caption{Error $\Delta d$ as a function of the sample size ($n_{tot}$)
for the standard Cantor set (A) and for the 1/2-Cantor set (B) with the different 
noise amplitudes, $\gamma = 10^{-1}, 10^{-2},10^{-3},10^{-4},10^{-5}, 0$.
The errors are denoted by the triangle-up triangle-down, diamond, circle, star and cross symbols, respectively. 
In the last cases, the power-like fits are added (solid lines).}
\end{center}
\end{figure}

From Fig.~1, it is clear that there is no power-like scaling of the error 
$\Delta d$ as a function of the noise amplitude ($\gamma$), 
especially for smaller data samples ($ \lesssim 10^3$). 
Also, the value of the error for $\gamma = 0.01$ 
({\it i.e.}, the relative noise as small as about $1$\%)
is rather large, of order $0.1$ and the relative value of the error 
is close to $20$\%. 
A surprisingly large error induced by a small additive noise was also found 
for the fractal stochastic point processes (FSPP). For $10^6$ data points 
the fractal dimension was estimated within $\pm 0.1$ accuracy, {\it i.e.},
with the error higher than $10$\% \cite{Lowen95}. 
Hence, the noise effect on the computational error is enhanced by 
at least one order of magnitude.

%%%%%%%%%%%%%%%%%%%%%%%%%fig%3%%%%%%%%%%%%%%%%%%%%%%%%%%%%%%%%%%%%%%%%%%%%%%%
\begin{figure}
\begin{center}
 \includegraphics[width=12.0cm,angle=0]{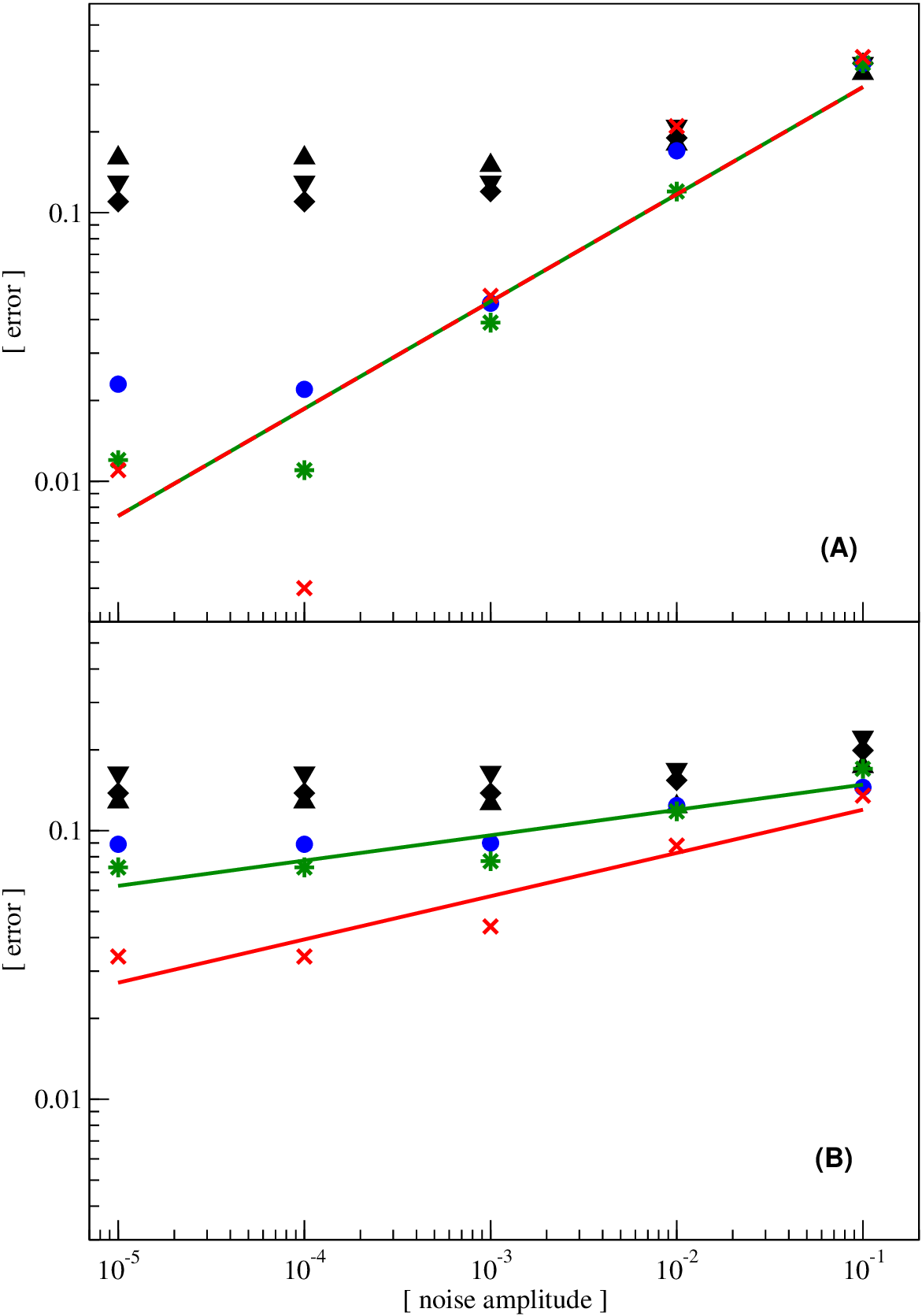}
 \caption{Error $\Delta d$ as a function of the noise amplitude ($\gamma$)
for the Sierpi\'nski triangle (A) and for the Sierpi\'nski carpet (B) with the different 
sizes of the data sample, $n_{tot} = 2^{10}, 2^{12}, 2^{14}, 2^{16}, 2^{18}, 2^{20}$.
The errors are denoted by the triangle-up triangle-down, diamond, circle, star 
and cross symbols, respectively. 
In the last cases, the power-like fits are added (solid lines).}
\end{center}
\end{figure}

In Fig~2., the error value is shown versus the sample size ($n_{tot}$). 
The different noise amplitudes,  
$\gamma = 10^{-1}, 10^{-2}, 10^{-3}, 10^{-4}, 10^{-5}, 0$, 
are denoted by the triangle-up, triangle-down, diamond, circle, star,
and cross symbols. 
The power fits for the smallest-noise amplitudes were also given for comparison. 
For the both fractals, the error value is almost independent of the sample size ---
it is dominated by the noise amplitude. 
Hence, by increasing the sample size one cannot improve numerical 
results. The triangle and diamond symbols are arranged almost horizontally. 
For a quite moderate noise 
(already below $1$\%, $\gamma=0.01\div 0.001$ --- triangles and diamonds), 
the error approaches the bound (\ref{errorbound}). 
Moreover, for a very small noise ($\gamma=10^{-5}$), the errors 
are up to three orders of magnitude bigger than the noise itself.

%%%%%%%%%%%%%%%%%%%%%%%%%fig%4%%%%%%%%%%%%%%%%%%%%%%%%%%%%%%%%%%%%%%%%%%%%%%%
\begin{figure}
\begin{center}
 \includegraphics[width=12.0cm,angle=0]{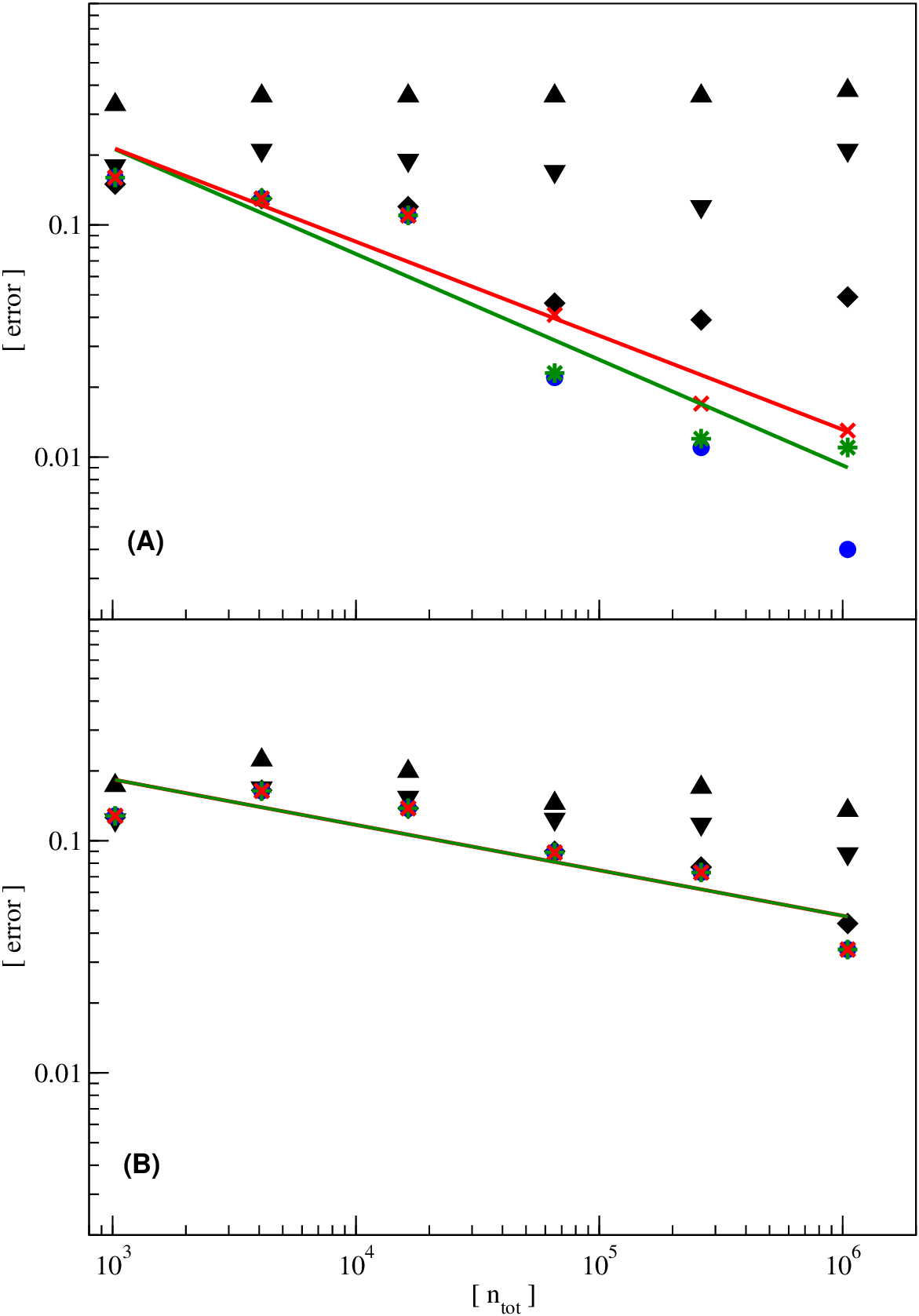}
 \caption{Error $\Delta d$ as a function of the sample size ($n_{tot}$)
for the Sierpi\'nski triangle (A) and for the Sierpi\'nski carpet (B) 
with the different noise amplitudes, $\gamma = 10^{-1}, 10^{-2},10^{-3},10^{-4},10^{-5}, 0$.
The errors are denoted by the triangle-up triangle-down, diamond, circle, star and cross symbols, respectively. 
In last two cases, the power-like fits are added (solid lines).}
\end{center}
\end{figure}

%%%%%%%%%%%%%%%%%%%%%%%%%%%%%%%% KONIEC POPRAWIONEJ CZESCI %%%%%%%%%%%%%%%%%%%%%%%%%%%%%%%%%%%%%%

It is interesting to check whether the 2-D fractals are also so sensitive
to an external noise. To this end, a similar analysis was performed for 
the Sierpi\'nski triangle (Figs.~3A,~4A) and for the Sierpi\'nski carpet 
(Figs.~3B,~4B). 
The fractal sets embedded in more dimensions are computationally more
demanding. It have been estimated that in order to have a comparable accuracy 
for the fractals 
embedded in 2-D space, the number of data points should be squared, in comparison 
to the 1-D fractals \cite{AZGczI}. 
Hence, the 2-D fractal sample sizes were taken in the range 
$n_{tot} = 2^{10}\div 2^{20}$.

In Fig.~3, the error dependence on the noise amplitude ($\gamma$) is shown 
for the Sierpi\'nski triangle (A) and for the Sierpi\'nski carpet (B). 
The different sample sizes, $n_{tot} = 2^{10}, 2^{12}, 2^{14}, 2^{16}, 2^{18}, 2^{20}$,
are represented by the triangle-up, triangle-down, diamond, circle, star and cross 
symbols, respectively. 
Again, an approximate power-like scaling is not observed. 
The total error ({\it i.e.} the sum of statistical and systematic errors,
$\Delta d$) was found 
in the range quite similar as for the 1-D fractals (Cantor sets). 
One can conclude that higher dimensional
fractals are not more sensitive to an additive external noise than 
the 1-D fractals. 
However, the error remains considerably large, at least about 
$10$\% for a small ($1$\% and less) noise and relatively large data samples
(of order $10^5$). 
The error (3) for a small noise ($\gamma\approx 10^{-5}$) 
is about three--four orders of magnitude larger (approaching $10^{-1}$).

 It is worth to mention that, in general, the external noise
{\it increases} the calculated value of the fractal exponents. 
This is intuitively clear, as the white noise embedded in the $E$-dimensional
space has the BC dimension equal to $E$. 
Hence, the obtained exponents for noisy fractals are usually 
overestimated and for large enough values of $\gamma$ one gets results
close to $E$. This effect prohibits the power-like scaling
of $\Delta d$ as a function of $\gamma$. 
In addition, one reaches natural cut-off for larger noise values, 
$\Delta d \le E$, where $E$ is the embedding dimension of the fractal.

\section{Summary and conclusions}

 In this paper the influence of an external additive 
noise with the amplitude $\gamma$ on the BC-algorithm calculations 
of the fractal exponents, for the fractal sets of size $n_{tot}$, was analysed. 
The numerical results are shown in Figs.~1--4. 

It has been found that the noise effect is surprisingly large --- 
a relativly tiny external noise ($\gamma=10^{-5}$) implies the 
error values (3) up to three--four orders of magnitude larger. 
A small noise is usually a part of any real data under investigation. 
Hence, one should be very careful when drawing conclusions
from numerically calculated fractal exponents for experimental data. 

As the error value is trivially limited by Eq. (4), for large 
noise amplitudes, the noise enhancement effect is saturated
and a power-like scaling of this effect does not exist. 
A strong sensitivity to the noise is in agreement with the results 
obtained earlier for FSPP \cite{Lowen95}. 
It is intuitively clear that the noise {\it increases} the computed fractal 
exponents, with the upper boumd $\Delta d \le E$. 
Moreover, as can be seen from  Figs.~2~and~4, the noise effect 
cannot be cured by taking much larger data samples.  Already for the
noise level approaching $0.1$\%, the sample size becomes irrelevant.

 In conclusion, one should treat very carefully the numerical results
of fractal exponents calculations for the data sets containing noise. 
This is very important since a small addition of noise is usually present 
in a majority of practical applications.

\end{document}